\input{aipcheck}

\documentclass[
    ,final            % use final for the camera ready runs
  ]
  {aipproc}

\layoutstyle{6x9}

\begin{document}

\title{Jets from massive protostars as gamma-ray sources: the case of 
IRAS~18162-2048}

\classification{97.10.Bt, 97.21.+a}
\keywords{Radiative processes: non-thermal-Stars: individual: 
IRAS 18162-2048}

\author{Anabella T. Araudo}{
  address={Centro de Radioastronom\'{\i}a y Astrof\'{\i}sica, 
Universidad Nacional Aut\'onoma de M\'exico, A.P. 3-72 (Xangari), 
58089 Morelia,  Michoac\'an, M\'exico}
}
\author{Luis F. Rodr\'{\i}guez}{
%  address={<common address for author2 and author3>}
}

\begin{abstract}
Protostellar jets are present in the later stages of the stellar formation.
Non-thermal radio emission has been detected 
from the jets and hot spots of some massive protostars, indicating the
presence of relativistic electrons there.  We are interested in  exploring if
these  non-thermal particles can emit also  at $\gamma$-rays.
In the present contribution we model the
non-thermal emission produced in the jets associated with the massive 
protostar  IRAS~18162-2048.  
We obtain that the $\gamma$-ray emission produced in this source is
detectable by the current facilities in the GeV domain. 
\end{abstract}

\maketitle

%%%%%%%%%%%%%%%%%%%%%%%%%%%%%%%%%%%%%%%%%%%%
%% MAINMATTER
%%%%%%%%%%%%%%%%%%%%%%%%%%%%%%%%%%%%%%%%%%%%

\section{Introduction}

 Massive stars are formed  within dense  molecular clouds,
accreting matter   onto the central protostar 
with the formation of a circumstellar disc and  forming bipolar 
jets and molecular  outflows. 
%In the case of jets from massive protostars, they are rare to detect because 
%massive stars forms  fast and in a very dense environment, and also they are 
%far away from us. 
In the last decade a handful of jets that 
emanate from massive protostars have been detected at radio wavelengths. 
In most of the cases, this radio 
emission has a negative spectral index ($\alpha$), indicating the non-thermal 
nature of 
it. However, only in the case of the jet associated with the source 
IRAS~18162$-$2048, polarization of radio 
emission has been measured [1] confirming that the radiation is produced by the 
synchrotron process. 

%Detection of synchrotron emission is evidence that there is a population of
%relativistic electrons in the source. Then, we are interested in explore if
%this population of non-thermal particles can produce radiation at gamma-rays.

\subsection{The jets of the source IRAS~18162-2048}

The famous Herbig-Haro (HH) objects called HH80-81 are the south component of
a system of several radio sources, located at a distance of $\sim$ 1.7 kpc.  
The central source has been identified with the luminous 
($L_{\star} = 1.7\times 10^4 L_{\odot}$) protostar IRAS~18162$-$2048, and HH80-N
is the northern counterpart of HH80$-$81.
The velocity of the jet close to the protostar has been estimated to be
as large as $v_{\rm j} \sim 1400$~km~s$^{-1}$ [2], 
whereas  HH~80 and HH~81 (located at $z \sim 2.5$~pc from IRAS~18162-2048) 
are moving with  
slower velocity, $\sim 150$~km~s$^{-1}$, and   HH80-N appears to be at rest.
Radio observations [2] show that the central source has a spectral index 
$\alpha \sim 0.1$, typical of free-free emission, whereas
HH80$-$81 and HH80-N are likely non-thermal sources, with 
$\alpha \sim -0.3$. Between the central protostar and the HH-objects there
is a chain of bright thermal and non-thermal knots. In particular, at
a distance $z_{\rm p} = 0.5$~pc from the central protostar, an elongated  
structure of polarized radio emission has been recently detected
through  observations carried out with
the  Very Large Array (VLA) [1].
% show that the jet of IRAS~18162-2048 is polarized.  
%This is the first protostellar jet with clear evidence of
%synchrotron emission. 
The detection of synchrotron radiation is evidence that there is 
a population of relativistic electrons in the source. 
These relativistic particles in the complex environment of
the massive molecular cloud where the protostar is being formed can
produce high-energy radiation through a variety of processes. 
In the present contribution we model the non-thermal
emission produced in the polarized jet of IRAS~18162-2048, from radio to 
$\gamma$-rays,  at
a distance $z_{\rm p} = 0.5$~pc from the central protostar, where the 
synchrotron emission is detected.

Radio observations suggest that the jets of 
IRAS~18162$-$2048 are precessing, because the direction of them
at $z_{\rm p}$ 
is different than the direction where the HH objects are located. 
We assume that now the head of the jet is at $z_{\rm p}$, and HH80-81 and
HH80~N are old ejections, $\sim$ 4000~yr before. 
%allowing to derive a dynamical age for the system similar to 4000~yr. 
%However, HH80-N appear to be at rest indicating that the density of 
%the molecular cloud in the north is larger than in the south,
%in the outkirts of the molecular cloud.  
The head of the jet moving through the molecular cloud forms  two shocks: a
 reverse shock in the jet, and a bow shock in the cloud, with velocities
$v_{\rm rs}|_{\rm bs} = v_{\rm j} - 3 v_{\rm bs}/4$ and
$v_{\rm bs} = v_{\rm j}/(1 + \sqrt{1/\chi})$, respectively, where
$\chi \equiv n_{\rm j}/n_{\rm mc}$ and $n_{\rm j}$ and $n_{\rm mc}$ are
the  densities of the jet  and the molecular cloud, respectively.
At $z_{\rm p}$ the width of the jet is $R_{\rm j} \sim 10^{17}$~cm, 
given $n_{\rm j} = \dot M/(\pi R_{\rm j}^2 v_{\rm j})\sim 2.6\times10^2$~cm$^{-3}$, 
where  $\dot M\sim 10^{-5}$~M$_{\odot}$~yr$^{-1}$ [3].   
%(This is the density of ionized matter if the matter that produce the free-free
%emission at the jet base does not recombines from the base up to $\sim 0.5$~pc,
%however matter can be recombined, on even ionized at a larger fraction in 
%the postshock region. We  neglect these efects in the present calculations)
At the location of HH80-N 
$n_{\rm mc}(z \sim 2.5\,{\rm pc}) \sim 400$~cm$^{-3}$ [4] is deduced
from $H_{\alpha}$ luminosity, and using that $n_{\rm mc} \propto z^{-1.5}$ [5] 
we obtain that at 0.5~pc  $n_{\rm mc} \sim 3\times10^{3}$~cm$^{-3}$. 
Thus we obtain that $\chi(z_{\rm p}) \sim 0.02$ giving an adiabatic 
reverse shock with velocity 
$v_{\rm rs} \sim v_{\rm j}$ and a radiative bow-shock.

\section{Particle acceleration and non-thermal emission}

Being that adiabatic shocks are propitious to accelerate particles 
via the Fermi-I mechanism, we assume that the electrons ($e$) that emit the 
detected synchrotron radiation at $z_{\rm p}$ are accelerated in the jet 
reverse shock (located also at $z_{\rm p}$). 
Even though there is no observational evidence of the presence of relativistic
protons ($p$) in the jet of IRAS~18162-2048, these particles can be 
accelerated via Fermi-I as well as electrons, and we will take into account
a population of non-thermal protons in our calculations. 
Particles accelerated in the reverse shock are  injected in the downstream 
region following a power-law energy distribution $Q_{e,p} \propto E_{e,p}^{-2}$.
The injection function $Q$ is modified by radiative and escape losses
in such a way that the part
of the electron spectrum $N_e$ that produces the synchrotron emission between 
1.5 and 15~GHz is $N_e \propto E_e^{-2}$. 
Considering equipartition between the magnetic and the relativistic 
particle energy densities, i.e. $B^2/(8\pi) = U_e + U_p$,  and the synchrotron 
fluxes given in [2], the magnetic field at $z_{\rm p}$ results 
$B \sim 0.1$~mG [1]. 

Electrons lose energy via synchrotron, Inverse Compton (IC) scattering and 
relativistic Bremmstrahlung, but escape losses via downstream matter advection 
is more efficient, with a timescale 
$t_{\rm adv} \sim 4\,R_{\rm j}/v_{\rm j} \sim 4\times10^9(v_{\rm j}/(1000 \,{\rm km\,s^{-1}}))^{-1}$~s as is shown in Figure~1 (left). 
Relativistic Bremsstrahlung losses have a timescale 
$t_{\rm Brem} \sim 1.4\times10^{11}(n_{\rm j}/260\,{\rm cm^{-3}})^{-1}$~s, being
this process the most efficient leptonic radiative channel at high energies.
IC losses are not relevant because at $z_{\rm p}$ the energy density of 
stellar photons is low: 
$U_{\rm ph} \sim L_{\star}/(\pi z_{\rm p}^2 c) \sim 9.4\times10^{-11}$~erg~s$^{-1}$.
Relativistic protons radiate through proton-proton 
($pp$) collisions on a timescale 
$t_{\rm pp} \sim 1.9\times10^{11}(n_{\rm j}/260\,{\rm cm^{-3}})^{-1}$~s.
Advection losses constrain the
maximum energy of electrons and protons given 
$E_{e,\rm max} = E_{p,\rm max} \sim 2.8$~TeV, and $N_{e,p} \sim Q_{e,p} \, t_{\rm adv}$.

\begin{figure}
\includegraphics[height=.35\textheight, angle=-90]{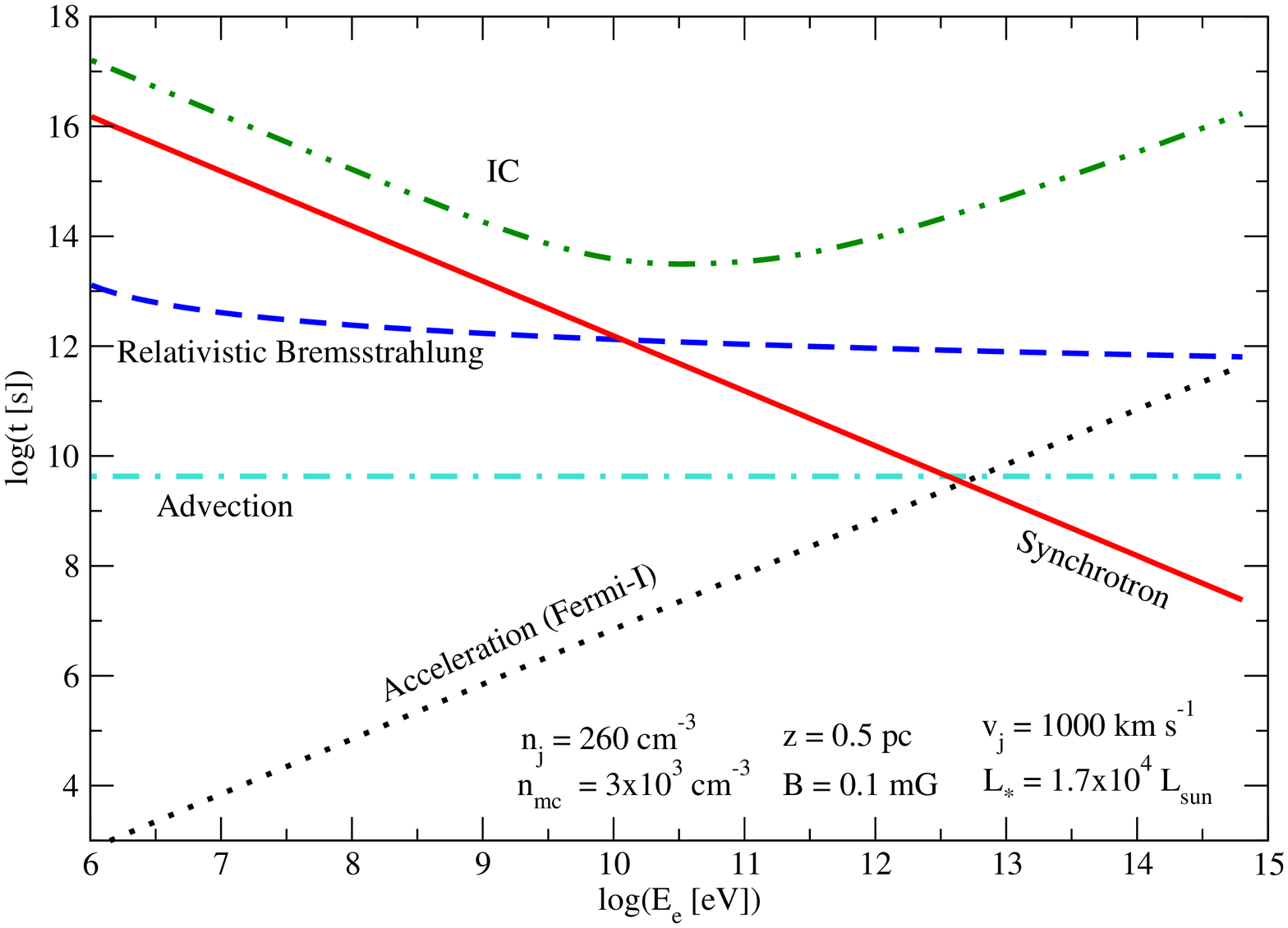}
\includegraphics[height=.35\textheight, angle=-90]{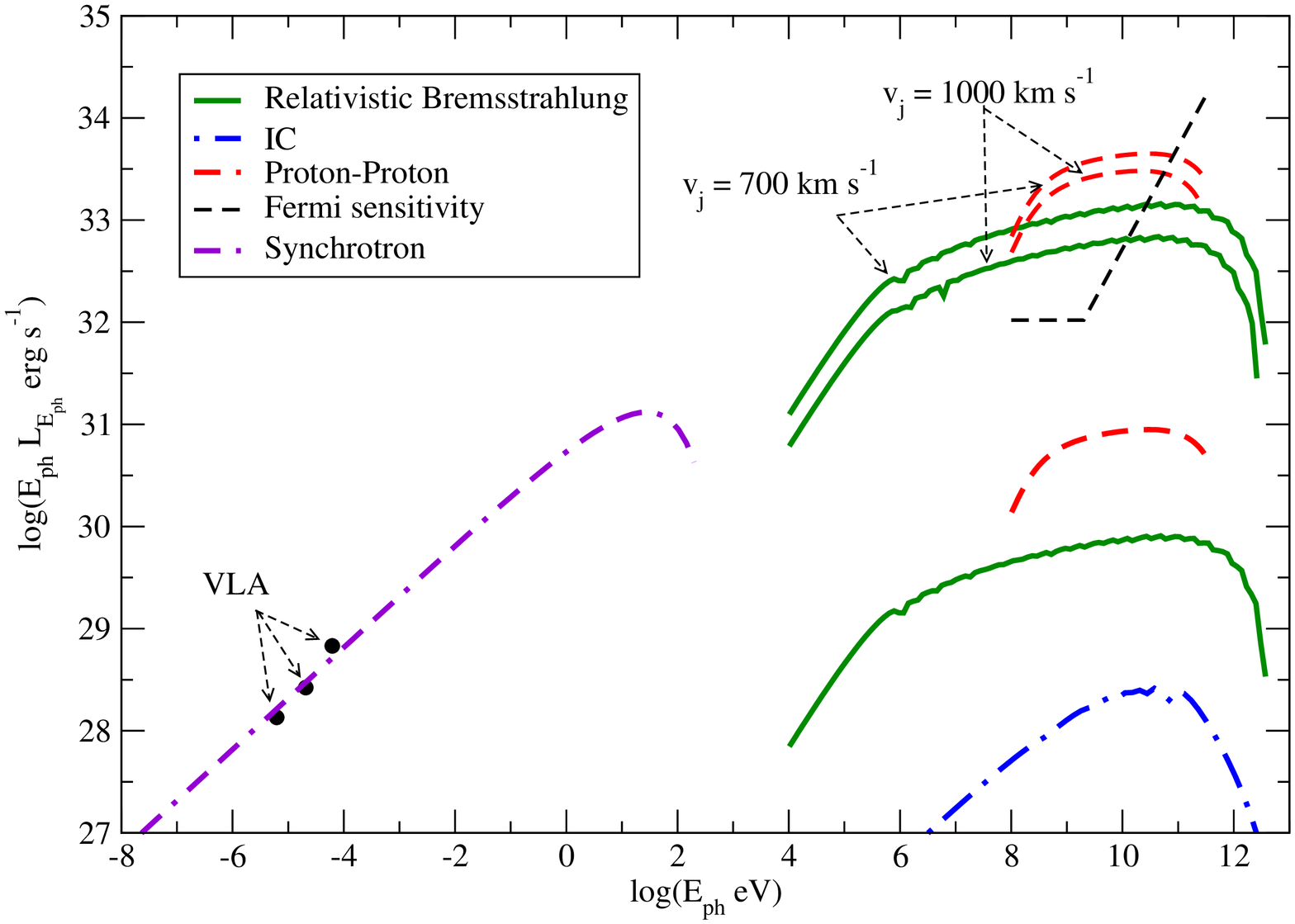}
\caption{\emph{Left}: Acceleration, escape, and electron cooling timescales.
The parameters of the jet and the molecular cloud used are specified in the 
plot. \emph{Right}: Spectral energy distribution. Relativistic Bremsstrahlung 
and $pp$ collisions are computed considering the shocked density of the jet
($4 n_{\rm j} = 1.1\times10^{3}$~cm$^{-3}$),
and also the shocked density of the molecular cloud 
($n_{\rm smc} = 9\times10^{5}$ and $1.8\times10^{6}$~cm$^{-3}$,
for $v_{\rm j} = 700$, and 1000~km~s$^{-1}$, respectively).}
\end{figure}

\subsection{Gamma-ray emission}

Knowing $N_e$ and $N_p$ we compute the spectral energy distribution 
considering  the radiative processes mentioned in the previous Section
(see Fig~1-right). We consider auto-absorption in the synchrotron spectrum 
and photon-photon attenuation at high energies, however this latter 
absorption process is not relevant at $\sim 0.5$~pc from the protostar.
Considering two different values of $v_{\rm j}$: 700 and 1000~km~s$^{-1}$, and
the jet shocked density as target of relativistic Bremsstrahlung, and $pp$,
the achieved bolometric luminosities of synchrotron radiation, IC scattering,
relativistic Bremsstrahlung, and $pp$ collisions result 
$L_{\rm sync} \sim 5\times10^{31}$,
$L_{\rm ic} \sim 2\times10^{29}$,   $L_{\rm Brem} \sim 8\times10^{30}$, and
$L_{\rm pp} \sim 5\times10^{31}$~erg~s$^{-1}$, respectively. We note that
in this case
$v_{\rm j}$ only affect the values of  $E_{e,\rm max}$ and $E_{p,\rm max}$, given
very similar bolometric luminosities.

We have mentioned that the most important cooling channels for electrons and
protons are relativistic Bremsstrahlung and $pp$ collisions, respectively.
However, specific luminosities not larger than $\sim 10^{30}$~erg~s$^{-1}$ 
(relativistic Bremsstrahlung) and $10^{31}$~erg~s$^{-1}$ ($pp$)
are achieved in the GeV domain by these radiation mechanisms.  
In order to produce a large amount of $\gamma$-rays, an increment in the
density of target particles is needed. This can occur if we consider 
that, being the large densitiy contrast between $n_{\rm j}$ and matter
downstream the bow shock, both materials can be mixed by  
Rayleigh-Taylor (RT) instabilities. 

The radiative bow shock in the molecular cloud compresses the matter 
up to a density $n_{\rm smc} \sim 10^5 (n_{\rm mc}/10^3\,{\rm cm^{-3}})(v_{\rm bs}/100\,{\rm km~s^{-1}})^2$~cm$^{-3}$ after cooling via thermal emission up to 
reach a temperature of $10^4$~K.
Thus, being the large density contrast between $4\,n_{\rm j} \sim 10^4$~cm$^{-3}$
and $n_{\rm smc}$, 
RT instabilities grow up to a wavelength $1/R_{\rm j}$ on a timescale 
$t_{\rm RT} \sim 0.1 t_{\rm j}$, where $t_{\rm j} \sim z_{\rm p}/v_{\rm j}$ [6,7]. 
Thus, 
at $z_{\rm p}$ the emitter (i.e. the reverse-shock downstream region) 
will be denser than $4\,n_{\rm j}$, and reaching a maximum value of
$n_{\rm smc}$. Considering 
the maximum increase in the density of the emitter, i.e. efficient RT mixing,
detectable values of  $\gamma$-rays in the GeV domain are achieved,
with specific relativistic Bremsstrahlung and $pp$ luminosities larger than
$\sim 5\times10^{32}$ and $5\times10^{33}$~erg~s$^{-1}$, respectively, 
as is shown in Fig~1 (right). 
However, note that detectable  emission
is achieved even in the case of less efficient mixing. By increasing the 
density of the emitter $\sim 10$ is enough to obtain detectable values of 
$\gamma$-rays. 

\section{Conclusions}

The massive protostar IRAS~18162-2048 presents very collimated and powerful 
jets, and  an accretion disc, characteristics that put this source
as the prototype of a massive star formation system. In addition to that, 
the jet of IRAS~18162-2048 has a large velocity ($\sim 1000$~km~s$^{-1}$) 
and it is  the unique protostellar jet with detection of polarized radio 
emission, mapping the 
direction of the magnetic field parallel to the jet velocity.
The large value of $v_{\rm j}$ and the detection of polarization at radio
wavelengths
make the jets of IRAS~18162-2048  potential sources of $\gamma$-rays [6]. 

In the present contribution we model the non-thermal emission (from radio
to $\gamma$-rays) produced at the
location of the jet where synchrotron emission is clearly detected. 
Constraining the model with VLA data, 
%and retaining only two free parameters ($v_{\rm j}$ and $$), 
and considering RT mixing between jet and molecular cloud
shocked matter we obtain that the $\gamma$-ray emission produced in  the 
north jet of IRAS~18162-2048 can be detected by the \emph{Fermi} 
satellite in the GeV domain. We note that besides  IRAS~18162-2048, there
are other massive protostars that can be also $\gamma$-ray sources, as 
the case of IRAS~16547-4247 [8].
%a more sophisticated model 
%of the $\gamma$-ray emission from protostellar jets will be presented in a 
%following paper. 
If high energy radiation from massive protostellar jets is detected,  
$\gamma$-ray astronomy can be used to 
shed light on the star forming process, and cosmic ray
acceleration inside molecular clouds.

%%%%%%%%%%%%%%%%%%%%%%%%%%%%%%%%%%%%%%%%%%%%%%%%%%%%%%%%%%%%%%%%%%%%%%%%%%%
\begin{theacknowledgments}
This work is supported by CONACyT, Mexico and PAPIIT, UNAM.
\end{theacknowledgments}

\bibliographystyle{aipproc}   % if natbib is available

\end{document}